\documentstyle[twoside,fleqn,espcrc2,amsmath,epsfig,bm]{article}

\title{Pomeron-Odderon interference effects in
 electroproduction of $\pi^+\,\pi^-$\thanks{Talk given by L.Sz. at the 9th
International
High-Energy Physics Conference QCD '02, Montpellier, France, 2-9th July 2002.}
\thanks{Work supported in part by the TMR and IHRP Programmes of the
European Union, Contracts No.~FMRX-CT98-0194 and No.~HPRN-CT-2000-00130.}}

\author{Ph. H\"agler\address{CTP, MIT, Cambridge, MA 02139, USA and  Universit{\"a}t
   Regensburg, D-93040 Regensburg, Germany},
B. Pire\address{CPhT, {\'E}cole Polytechnique, F-91128 Palaiseau, France}%
\thanks{Unit{\'e} mixte C7644 du CNRS.},
L. Szymanowski\address{CPhT, {\'E}cole Polytechnique, France and Soltan
Institute for Nuclear Studies, Warsaw,
Poland}
and
O.V.~Teryaev\address{Bogoliubov Lab. of Theoretical Physics, JINR, 141980 Dubna,
Russia}}

\begin{document}

\begin{abstract}
We review the results of our studies on the charge and the spin asymmetries in the
electroproduction of $\pi^+\,\pi^-$ which are the observables sensitive to the
Pomeron-Odderon interference. 
\end{abstract}

\maketitle

\section{Introduction}

The description of the Odderon \cite{LN} within QCD \cite{BKP} was in recent years 
the subject of
several
studies \cite{Levodd,JW,Vacca1,Korch}. Being a partner of the Pomeron
\cite{BFKL} which carries
the quantum numbers of the vacuum and which is well known from the study of
 diffractive
processes, the odderon still remains a mistery. Although it differs from the Pomeron
only by its charge conjugation property and from the point of view of general
principles one could expect that its exchange  leads to effects of a comparable
magnitude to those with the Pomeron exchange, the Odderon still escapes 
experimental verification. The QCD predictions lead to a rather small  cross
section for the diffractive $\eta_c-$production \cite{KM,Engel,Vacca2}.
Recent experimental studies of exclusive $\pi^0-$production at HERA 
\cite{Olsson}
show also a very small cross section for this process, which stays in contradiction
with theoretical predictions based on the stochastic vacuum model 
\cite{Dosh,Doshrecent}, see also \cite{Schatz}.

Because of these difficulties a new approach to Odderon search is required. In
Ref.~\cite{Brodsky} it was proposed to search for  Odderon effects by studying the
charge asymmetries in  open charm production, which are linear in the Odderon
scattering amplitude. These ideas were further developed in \cite{Nikolaev} where it
was proposed to study odderon effects in the soft photoproduction of two pions.

In two  recent papers \cite{HPST,HPSTlong} we proposed to
  study   the diffractive electroproduction
of a $\pi^+ \,\pi^-$ pair
to search for the QCD-Odderon at the amplitude level.
The main difference of our studies of the electroproduction
process with respect to Refs. \cite{Brodsky,Nikolaev} is to work in 
a perturbative framework
which we believe enables us to
derive more founded predictions in an accessible kinematical domain.

\section{The theoretical framework}

We study the electroproduction of $\pi^+\,\pi^-$ pair which
 proceeds through a virtual photon-proton reaction  
\begin{equation}  
\label{gp}
\gamma^* (q, \epsilon)\;\; N (p_N) \to \pi^+(p_+)\;\; \pi^-(p_-)\;\;
N^{\prime}(p_{N^{\prime}})\;.
\end{equation}

In the Born approximation the Pomeron exchange is described by two gluon exchange in
a colour singlet state (see Fig.~1) and the scattering amplitude in the
impact representation has the form of a convolution in the 2-dimensional transverse
momentum space

\begin{figure*}[htbp]
\epsfig{file=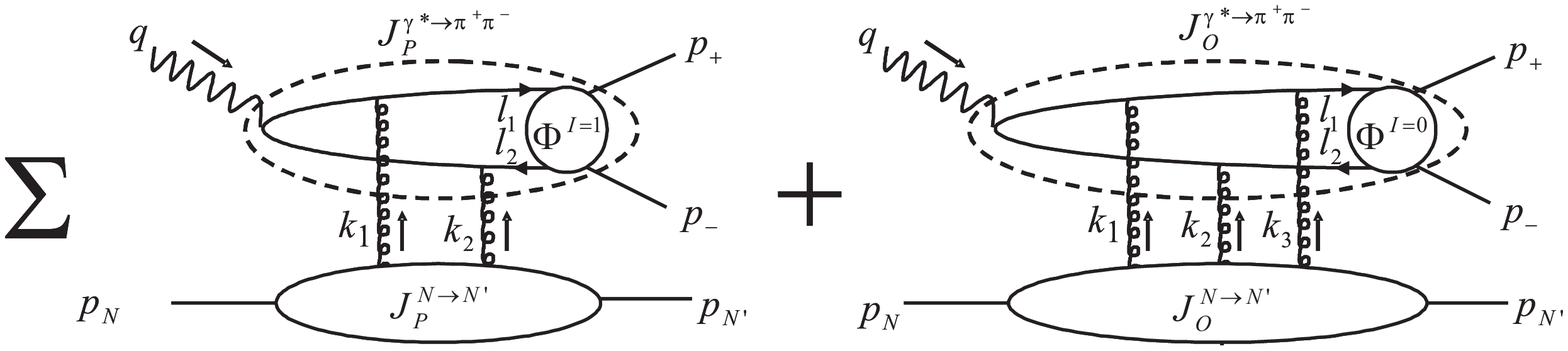,width=14cm}
\caption{{\protect\small Feynman diagrams describing $\pi^+ \pi^-$
electroproduction in the Born approximation }}
\label{Fig1} 
\end{figure*}
\[
{\cal M}_{L/T}(P) = -i\,s\,\int\;\frac{d^2 \vec{k}_1 \; d^2 \vec{k}_2 \;
\delta^{(2)}(\vec{k}_1 +\vec{k}_2-\vec{p}_{2\pi})}{(2\pi)^2\,\vec{k}_1^2\,%
\vec{k}_2^2} \nonumber
\]
\begin{equation}
J_P^{\gamma^* \rightarrow \pi^+\pi^-}(\vec{k}_1,\vec{k}_2)\cdot
J_P^{N \rightarrow N^{\prime}}(\vec{k}_1,\vec{k}_2)\;,
\label{pom}
\end{equation}
where $J_P^{\gamma^* \rightarrow \pi^+\pi^-}(\vec{k}_1,\vec{k}_2)$  and $%
J_P^{N \rightarrow N^{\prime}}(\vec{k}_1,\vec{k}_2)$ are the impact factors
for the transition $\gamma^*_{L/T} \to \pi^+\ \pi^-$ via Pomeron exchange  and
of the
nucleon in the initial state $N$ into the nucleon in the  final state
$N^{\prime}$.

The corresponding representation for the Odderon exchange, \emph{%
i.e.} the exchange of three gluons in a colour singlet state, is given by
the formula (see Fig.~1)  
\[
{\cal M}_{L/T}(O) =  -\frac{8\,\pi^2\,s}{3!}  \nonumber
\]
\[
\int\;\frac{d^2 \vec{k}_1 \; d^2 \vec{k%
}_2 d^2 \vec{k}_3\; \delta^{(2)}(\vec{k}_1 +\vec{k}_2 +\vec{k}_3-\vec{p}%
_{2\pi})}{(2\pi)^6\,\vec{k}_1^2\,\vec{k}_2^2\,\vec{k}_3^2} \nonumber 
\]
\begin{equation}
J_O^{\gamma^*
\rightarrow \pi^+\pi^-}\cdot J_O^{N \rightarrow N^{\prime}}\;.
\label{odd}
\end{equation}
$J_O^{\gamma^* \rightarrow \pi^+\pi^-}(\vec{k}_1,\vec{k}_2,\vec{k}_3)$
and $J_O^{N \rightarrow N^{\prime}}(\vec{k}_1,\vec{k}_2,\vec{k}_2)$ are the
corresponding impact factors for the same transitions  via Odderon
exchange.

The impact factors in Eqs. (\ref{pom}) and (\ref{odd}) involve  hard parts
calculated perturbatively by standard methods and nonperturbative  light-cone
generalized distribution amplitudes (GDA) \cite{DGPT} describing the transition of
a
$q\bar q$ pair
 into the $\pi^+\pi^-$ final state. 

The choice of appropriate two pion GDAs is a point of  crucial importance.
We parametrize them according to results of recent studies 
\cite{POL,DGP}. The GPD related to the Odderon exchange involves (see
\cite{HPSTlong} for
details) the
$s-$ and $d-$ partial waves contributions corresponding to $f_0$ and $f_2$ meson
poles in the
$\pi\pi$ elastic scattering.
The GPD coupled with Pomeron exchange involves a $p-$ partial wave described by
the
phase shift and Breit-Wigner formula for the $\rho-$meson.

The second important point of our approach is the choice of the impact
factor describing the Pomeron and the Odderon coupling to the target nucleon.
Here we are forced to use  phenomenological models derived to fit 
hadronic data \cite{protonP,protonO}.

\section{The charge asymmetry}

The charge asymmetry is defined as
\[
\label{ca}
A(Q^{2},t,m_{2\pi }^{2},y,\alpha )= \nonumber 
\]
\[
=\frac{\sum\limits_{\lambda =+,-}\int \cos
\theta
\,d\sigma (s,Q^{2},t,m_{2\pi }^{2},y,\theta , \alpha, \lambda )}{%
\sum\limits_{\lambda =+,-}\int d\sigma (s,Q^{2},t,m_{2\pi }^{2},y, \alpha,  \theta
,\lambda )}
\nonumber
\]
\begin{equation}
=\frac{\int d\cos
\theta \cos\theta \;N_{charge}}{\int d\cos
\theta \;D}
\end{equation}
where the differential cross section $d\sigma (s,Q^{2},t,m_{2\pi }^{2},y, \theta
,\lambda )$ is a function of the cms scattering energy square $s$, the photon
virtuality $Q^2$, the momentum transfer square $t$, the invariant mass of the pion
pair $m_{2\pi }$, the longitudinal momentum fraction of the initial electron
carried
by
the photon $y$, the polar decay angle (in the cms of the pion pair) $\theta$,
the
azimuthal angle $\alpha$ between the leptonic plane and the plane of reaction
(\ref{gp})
and of the helicity of the scattered electron $\lambda$. This asymmetry is a
measure of
the difference between
the number of events in which the $\pi^+$ is produced in forward hemisphere (in
the cms
of the pion pair) and in the backward hemisphere.


Introducing the reduced scattering amplitude ${\cal A}_T(P/O)$ for transversely
polarized photons:
${\cal M}_T(P/O) \equiv (\vec{\epsilon}(T)\cdot\vec{p}_{2\pi})\,{\cal A}_T(P/O)$
the numerator $N_{charge}$ and denominator $D$ of the asymmetry have the forms
\[
N_{charge} = 8(1-y)\mbox{Re}\left[ {\cal M}_{L}(P){\cal M}%
_{L}^{*}(O)\right] \nonumber
\]
\[
+4(2-y)\sqrt{1-y}|\vec{p}_{2\pi }|\cos \alpha \;\mbox{Re}\left[
{\cal A}_{T}(P)%
{\cal M}_{L}^{*}(O)\right. \nonumber
\]
\[
\left. +{\cal A}_{T}(O)\;\;{\cal M}_{L}^{*}(P)\right] +2(1+(1-y)^{2}+2(1-y)\cos
2\alpha ) \nonumber
\]
\begin{equation}
\hspace{-.7cm} |\vec{p}_{2\pi }|^{2}\mbox{Re}%
\left[ {\cal A}_{T}(P){\cal A}_{T}^{*}(O)\right]
\label{nomca}
\end{equation}
and

\[
\hspace{-.9cm}D =4(1-y)\left| {\cal M}_{L}(P)+{\cal M}_{L}(O)\right| ^{2}
\]
\[
+4(2-y)\sqrt{1-y}|\vec{p}_{2\pi }|\cos \alpha \;\mbox{Re}\left[ \left(
{\cal A}_{T}(P) +{\cal A}_{T}(O)\right)\right.
\]
\[
\left. \hspace{-.9cm}\left(
{\cal M}_{L}^{*}(P)+{\cal M}%
_{L}^{*}(O)\right) \right]
+(1+(1-y)^{2}
\]
\begin{equation}
\hspace{-.7cm}+2(1-y)\cos 2\alpha)|\vec{p}_{2\pi }|^{2}\left|
{\cal A}_{T}(P)+{\cal A}_{T}(O)\right| ^{2}
\label{denca}
\end{equation}

\begin{figure}[htbp]
\epsfig{file=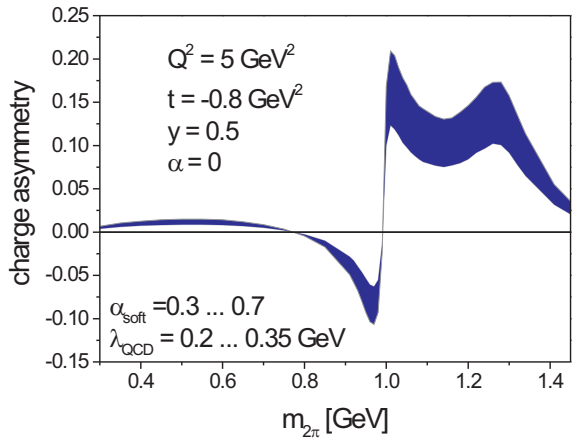,width=\columnwidth}
\caption{{\protect\small  Error bands resulting from a variation of  
$\Lambda_{QCD}$ and the soft coupling $\alpha_{soft}$. }}
\label{Fig2}
\end{figure}

As one example of our results we present in Fig.~2 a plot of the charge asymmetry
as a function of the two pion invariant mass $m_{2\pi}$ and for the  typical
for HERA values of the other parameters. The error band corresponds to a variation
of
$\Lambda_{QCD}$ and the soft coupling constant in the impact factor 
$J_{P/O}^{N \rightarrow N^{\prime}}$. The characteristic  $m_{2\pi}-$dependence of
the asymmetry is completely understood in terms of the $\pi \pi$ phase shifts
$\delta_0$, $\delta_1$, $\delta_2$ and its dependence on them $\sim \sin
(\delta_{0,2} - \delta_1)$. 

The important source of corrections to above prediction is due to the
effects of $K\,\bar K$ threshold in description of asymmetries for $m_{2\pi} >
1\,$GeV. The  example of their magnitude  is presented in Fig.~3
which shows the 
effect of taking into account the inelasticity factor $\eta$ \cite{Kloet} in the
scattering
amplitude. We see that  the charge asymmetry still remains sizable and it has a
characteristic pattern.

\section{The spin asymmetry}

The spin asymmetry related to the polarizations of incomming electron and it is
defined as
\[
\hspace{-.8cm}A_{S}(Q^{2},t,m_{2\pi }^{2},y )= \nonumber
\]
\[
\hspace{-.9cm}=\frac{\sum\limits_{\lambda
=+,-}\lambda \int \cos \theta \,d\sigma (s,Q^{2},t,m_{2\pi }^{2},y,
,\theta ,\lambda )}{\ \sum\limits_{\lambda =+,-}\int d\sigma
(s,Q^{2},t,m_{2\pi }^{2},y,\alpha ,\theta ,\lambda )}
\]
\begin{equation}
\hspace{-.9cm}=\frac{\int d\cos
\theta \cos\theta \;N_{spin}}{\int d\cos
\theta \;D} 
\mbox{}
\label{sa}
\end{equation}
with the numerator given by
\[
N_{spin}=4y\sqrt{1-y}\sin \alpha \;|\vec{p}_{2\pi }|\;\mbox{Im}\left[
{\cal M}_{L}(P){\cal A}%
_{T}^{*}(O) \right.
\]
\begin{equation}
\left.\hspace{-1cm}+{\cal M}_{L}(O){\cal A}_{T}^{*}(P)\right]
\label{nomsa}
\end{equation}
while $D$ is of course the same quantity as in the case of the charge asymmetry
(\ref{denca}). In Fig.~3 we show the plot of spin asymmetry. Comparing it with
Fig.~2 we conclude that the magnitude of spin asymmetry is quite small.

\begin{figure}[htbp]

\epsfig{file=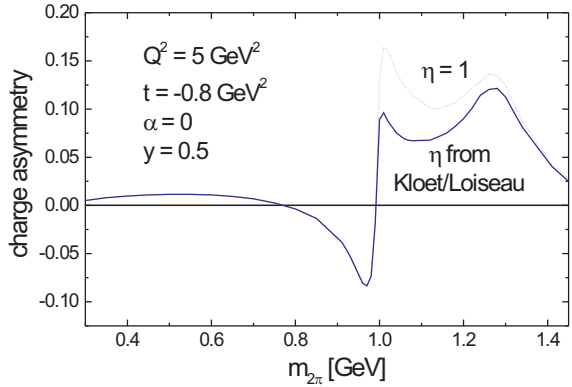,width=\columnwidth}
\vspace*{-1cm}
\caption{{\protect\small $m_{2\pi}-$dependence of the charge asymmetry
with
the inelasticity factor $\eta (m_{2\pi})$ (solid curve) and with $\eta
(m_{2\pi})=1$ (dotted curve) for $Q^2=5$GeV$^2$, $t=-0.8$GeV$^2$, $%
\alpha=0$, $y=0.5$  }}
\label{Fig4}
\end{figure}

\vskip.1in
\noindent {\bf Discussion}
\vskip.1in
\noindent {\bf P. Minkowski}, Bern: {\it Do you think that the asymmetries you study
could be
used for fixing a form of GDAs ?}

\noindent L. Szymanowski: The motivation of our work was rather reverse: we tried to
evaluate asymmetries assuming that we know  GDAs. Of course if the data on
asymmetries will be precise enough one could also think to do what you ask.

\begin{figure}[htbp]
\epsfig{file=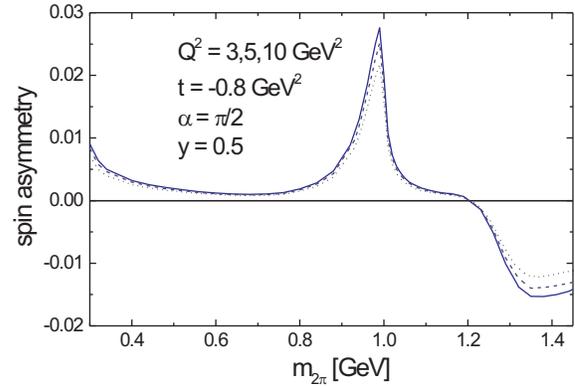,width=\columnwidth}
\caption{{\protect\small $m_{2\pi}-$dependence of the spin asymmetry at
$t=-0.8$
 GeV$^2$, $Q^2=3$ (solid line), 5 (dashed line), 10 (dotted line)  GeV$^2$  }}
\label{Fig3}
\end{figure}

\end{document}